\documentclass[a4paper, 11pt]{article}

\usepackage{amssymb}
\usepackage{amsmath}
\usepackage{latexsym}
\usepackage{graphicx}

\usepackage{url}
\usepackage{xcolor}
\definecolor{newcolor}{rgb}{.8,.349,.1}

\usepackage[citebordercolor=white]{hyperref}

\usepackage{algorithm}
\usepackage{algpseudocode}

\begin{document}

\title{Space weather dependence of decameter-scale ionospheric irregularities spatial distribution} 

\author{Oleg I.Berngardt, Ivan A.Lavygin\thanks{ISTP SB RAS, 664033, Lermontova Str, 126A, Irkutsk, Russia, email: berng@iszf.irk.ru}}

\maketitle

\begin{abstract}

This paper describes an approach to constructing nowcast and forecast predictive models
for the occurrence of high-latitude decameter ionospheric
irregularities based on SuperDARN and SECIRA radar data and space
weather parameters. The method relies on self-learning pattern recognition in the
dynamics of space weather parameters,
that influence the irregularities formation. The proposed
approach is a data-driven, multiscale method in which the spatial
distribution of irregularities is approximated using a neural network
of an original architecture, inspired by graph neural networks.

The model was trained and tested using data from 11 radars during
the period 2010-2025 (SuperDARN: CVW, ICE, ICW, INV, PGR, PYK, RKN,
SAS, STO; SECIRA: EKB, MAGW). It is shown that the model achieves
optimal performance when using geomagnetic coordinates MLT,MLAT. 
The optimal set of parameters for
the best nowcast and forecast over studied set of parameters 
is a combination of geomagnetic and solar disturbance
indices ($a_{p}$, $K_{p}$, $AE$, F10.7) from ground-based observations,
and satellite-measured solar wind and interplanetary magnetic field
parameters ($P_{sw}$, $E_{sw}$, $\widetilde{B_{z}}$, $T_{sw}$).
Hourly mean values for all the parameters were used in the analysis.
The results indicate that acceptable forecast quality is achieved
with a lead time of up to 7-9 hours, with the highest quality reached
for nowcast (zero lead time). It is shown that the parameters $\widetilde{B_{z}}$
and F10.7 are significant for nowcasting only and can be neglected for other
lead times. It is shown that $AE$ and $T_{sw}$ are significant for mid-term forecast 
(\textgreater 2hours ahead) and can be neglected for short-term predictions.

Statistical relationships between the maximum latitude of the equatorial
boundary of the irregularities and space weather parameters are also
discussed.

\end{abstract}

\makeatother

\section{Introduction}

The problem of modeling the spatio-temporal distribution of irregularities
and forecasting them based on measurable space weather parameters has stated for quite
long time \cite{Booker_1938}.

Models of spatial distribution of high-latitude small-scale irregularities
can be approximately categorized into empirical and physical ones.
Empirical models, in turn, can be subdivided into autoregressive models
(which use only recent observations of ionospheric irregularities to
forecast their future behavior) and regression models (which
use only observations of other parameters for forecasting irregularities). 
A review of many models can be found in \cite{Priyadarshi_2015}. Ionospheric
irregularities have complex dynamics dependent on external conditions,
and their physical modeling is discussed, for example, in 
\cite{Huba_2010,Deshpande_2014}.

Numerous models for the spatial distribution of irregularities
in the equatorial ionosphere have been developed. One early example
is the Scintillation Network Decision Aid (SCINDA) model for
irregularity regions at equatorial latitudes in the VHF (250 MHz)
frequency band. It uses data from spatially separated
ground-based receivers of geostationary satellite signals, as well
as solar (F10.7) and geomagnetic ($K_{p}$) activity data \cite{Groves_1997}.
Recent modifications of this model also exist \cite{Caton_2004}.

The physical processes controlling the formation
of small-scale irregularities in polar and equatorial regions
are often close \cite{Tsunoda_1988}, allowing approaches used in
equatorial ionosphere to be adapted for auroral one. For example, the
equatorial Wideband
Ionospheric Scintillation Model (WBMOD) based on a power-law fluctuation
spectrum \cite{Secan_1995} was subsequently modified for high latitudes
\cite{Secan_1997}. There are also studies of optimizing 
the parameters of this model using machine learning techniques \cite{Molina_2026}.

One of the global models to predict signal scintillation level (caused by small-scale irregularities presence)
in UHF radio bound is
the Global Ionospheric Propagation Model (GIM). It is based on the
NeQuick background ionospheric model and incorporates the spectrum
of irregularities estimated from global navigation satellite system (GNSS) network observations, 
taking into account NeQuick input parameters - local time, season, and solar activity 
(as F10.7 index) \cite{B_niguel_2011}. The model can also be modified to predict
scintillation maps in UHF band in real time \cite{Hamel_2014}.

A problem of irregularities prediction in the HF band is the lack of 
a dense observational network.
Today the regular observations are made by vertical
and oblique sounding ionosondes and HF radars of the SuperDARN\cite{Nishitani2019}, CN-DARN\cite{CNDARN},
and SECIRA\cite{SECIRA} networks.
In practice, using small observational networks forces to the use of low-parameter models to describe
the characteristics of regions containing irregularities. One widely
used approach involves spherical harmonic approximation \cite{Ruohoniemi_1996,Shepherd_2000,Cousins_2010,Mestici_2026}.
A dynamic model \cite{Mestici_2026}, uses data from the Swarm
satellite mission, and based on the interplanetary magnetic
field (IMF) orientation (categorized into six IMF clock angle sectors),
three levels of solar activity, and three seasons. Due to one of the
physical driver for the formation of small-scale high-latitude
irregularities is the electric field, 
the modeling strategy could involve high-latitude convection models built 
from SuperDARN radar data and based on IMF clock angles
\cite{Ruohoniemi_1996,Shepherd_2000}. The convection model was improved
in \cite{Cousins_2010} by including the interplanetary
electric field ($E_{sw}$) as a model parameter. 
Recent papers try to improve this
model using neural networks \cite{Gao_2025,Gao_2026}. Another approach
use regression-based multi-scale model \cite{Zhao_2026} to predict the irregularities fluctuation 
level at each measurement location independently.

The key problems in constructing a model of the spatiotemporal distribution
of irregularities in the HF range can be formulated as follows.

Problem 1: selection of model input parameters. To solve this problem,
the following approaches are typically used:
\begin{itemize}
\item an empirical approach: postulating input parameters based on physical
assumptions on the processes of irregularities formation 
\cite{Ruohoniemi_1996,Shepherd_2000,Cousins_2010,Gao_2025,Gao_2026};
\item a data-driven approach: selecting the maximum available number of
measured parameters and analyzing their importance \cite{McGranaghan_2018}.
\end{itemize}

Problem 2: taking into account the temporal dynamics
of parameters. Typically, the following approaches are used
to solve this problem:
\begin{itemize}
\item a regression-based approach, assuming that the parameter value directly
and nearly simultaneously determines the intensity of irregularities
\cite{Ruohoniemi_1996,Shepherd_2000,Cousins_2010,Gao_2025,Gao_2026};
\item a delay-based approach, in which the delay between input model parameter 
and irregularity variations (or temporal scale of input parameter averaging) 
is determined from the data to select the most significant ones
corresponding to the most important lag or averaging period \cite{McGranaghan_2018};
\item a multiscale approach, based on the fact that different
patterns in parameter temporal dynamics affect the predicted
parameter in different ways \cite{Zhao_2026}.
\end{itemize}

Problem 3: choosing a method for spatial approximation of the model
to areas where there are no observations. The following approaches are
used to solve this problem:

\begin{itemize}
\item spatial interpolation of the results, or their approximation using
spherical harmonics and searching for their coefficients \cite{Ruohoniemi_1996,Shepherd_2000,Cousins_2010};
\item training the model in such a way that spatial coordinates are the
input parameters of the model \cite{Gao_2025,Gao_2026} and the approximation
is performed by the resulting model itself;
\item predicting only in the locations where there were observations 
in training data \cite{Zhao_2026}.
\end{itemize}

This paper describes a predictive model
for the probability of observing the high-latitude ionospheric irregularities
based on SuperDARN and SECIRA radar data.
The proposed method is a data-driven, multiscale
approach where the spatial distribution of irregularities is approximated
by a neural network model optimally trained on available experimental
data.
The method uses pattern
recognition over the sequences of input parameters to improve
the prediction quality. The
input model parameters are found as a minimal set optimized for model
quality, observation coordinates are used as separate model inputs.

\section{Preliminary data analysis}

\subsection{Data used}

\begin{figure}
\includegraphics[scale=0.70]{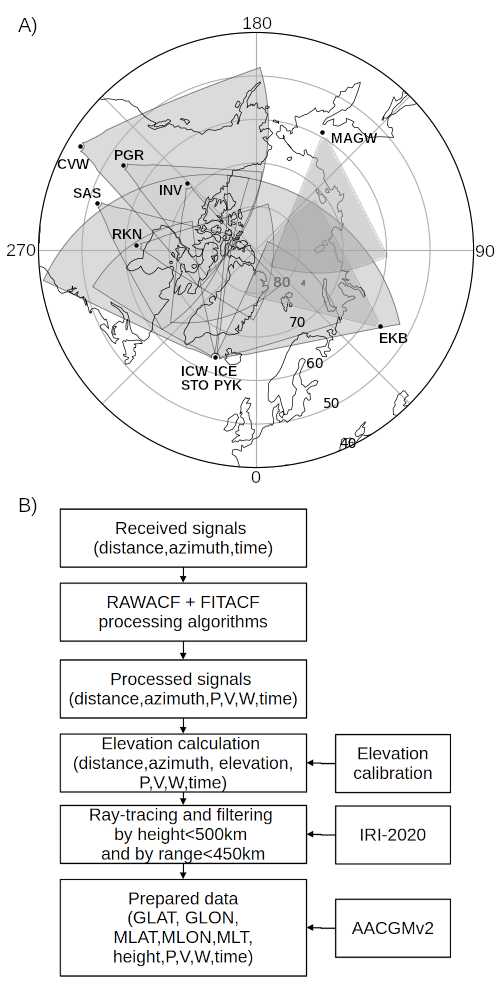}
\caption{(A) Radars fields-of-view in geographic coordinates; (B) Data preparation algorithm.}
\label{fig:RadarMap}
\end{figure}

To build the model, we used data from 11 SuperDARN and SECIRA radars
in the years of various solar and geomagnetic activity, shown in the Table \ref{tab:RadarsYears}. 
Their fields-of-view are shown in the Fig.\ref{fig:RadarMap}.
The data is divided to train over the period 2010-2023, 
and to validate over the period 2024-2025.

During training at least two years were selected for each radar - an year of relatively
high and an year of relatively low solar activity. The largest radar number corresponds to 2023.
Additionally, the
PYK radar with 8 years of data was used. This can be explained
by to two factors: it is necessary to build a model suitable for a
wide range of space weather parameters over sufficiently large
period of the solar cycle, and it is necessary to build
a model covering a large spatial range. Due to limited computing
power, such a solution seems reasonable. In the
future, it looks useful to apply the developed approach to a wider dataset.

\begin{table}
\caption{Radars used in the analysis, their geographical coordinates, and the years analyzed.
Data not used for model training are highlighted in bold.}

\begin{tabular}{|c|c|c|}
\hline 
Radar & Coordinates & Years\tabularnewline
\hline 
\hline 
CVW SuperDARN & 43N 120W & 2020,2023\tabularnewline
\hline 
EKB SECIRA & 56N 58E & 2021,\textbf{2025}\tabularnewline
\hline 
ICE SuperDARN & 64N 21W & \textbf{2024}\tabularnewline
\hline 
ICW SuperDARN & 64N 21W  & \textbf{2024}\tabularnewline
\hline 
INV SuperDARN & 68N 134W & 2018,2023\tabularnewline
\hline 
PGR SuperDARN & 54N 123W & 2020,2023\tabularnewline
\hline 
PYK SuperDARN & 64N 21W & 2010-2014,2016-2018\tabularnewline
\hline 
RKN SuperDARN & 63N 92W & 2018,2023\tabularnewline
\hline 
SAS SuperDARN & 52N 106W & 2020,2023\tabularnewline
\hline 
STO SuperDARN & 64N 21W & 2011,2014\tabularnewline
\hline 
MAGW SECIRA & 60N 150E & 2021,\textbf{2025}\tabularnewline
\hline 
\end{tabular}\label{tab:RadarsYears}
\end{table}

The radar elevation data was calibrated at meteor trails using the method described
in \cite{BERNGARDT_2026}. Signal propagation trajectories were calculated for
each powerful enough received signal using ray-tracing in the ionosphere described by IRI-2020 model\cite{IRI2020}.
After ray-tracing each found scattering points were located by geographic and geomagnetic coordinates (based on the
AACGMv2 model\cite{AACGMv2}, and by the height.

\subsection{Problem formulation}

Our primary objective is to find the dependence of the spatio-temporal
distribution of ionospheric irregularities (ionospheric scattering,
IS) observation probability on measureable space weather parameters.
Let us denote the vector of space weather parameters as $\vec{\theta}$
(later used as $\theta$). Let us denote the combination of space
weather parameters and spatial location coordinates $\vec{R}$ as
$\vec{\theta_{s}}=(\vec{\theta},\vec{R})$ (later used as $\theta_{s}$).
In the paper we will use standard notations $P(A)$ as probability of event
$A$, and $P(A|B)$ as conditional probability of event
$A$ given event $B$.

So our objective is to build a model describing IS observation
probability $P(IS|\theta_{s})$ under various conditions $\theta$
and in different locations $\vec{R}$ using data from coherent scatter
radars. In practice, however, we cannot directly estimate this probability
because we cannot detect the ionospheric scattering everywhere
it exists. Scattering detection using HF radars requires looking in
the optimal direction, selecting an appropriate frequency to ensure
the radiosignal trajectory meets the optimal scattering conditions,
having the radar operating at any given time, and other factors. Therefore
we need to formulate the problem of predicting $P(IS|\theta_{s})$
in a way that minimizes the effects of non-uniform measurement locations,
radar operation intervals, and changes in radar operating modes.

Each radar observes several types of signals (OBS): scattering from
the Earth's surface, meteor trail scattering, ionospheric scattering, and noise \cite{Nishitani2019}. The probability
of the radar network observing these signals under various conditions $\theta_{s}$ 
can be described by the probability $P(OBS|\theta_{s})$.
The relation between $P(OBS|\theta_{s})$ and $P(IS|\theta_{s})$
is defined by the Bayes' formula:

\begin{equation}
P(OBS|IS,\theta_{s})\cdot P(IS|\theta_{s})=P(IS|OBS,\theta_{s})\cdot P(OBS|\theta_{s})\label{eq:BaysesEq}
\end{equation}

The multipliers in this expression are interpreted as follows:

$P(IS|OBS,\theta_{s})$ is the probability that we are observing scattering
from ionospheric irregularities, if a signal has been received by
radars in space weather conditions and for spatial location given
by the vector $\theta_{s}$.

$P(OBS|IS,\theta_{s})$ - the probability of observing scattering
from ionospheric irregularities, if they exist in space weather conditions
and for spatial location given by the vector $\theta_{s}$;

The most difficult factor in the formula is to determine experimentally
$P(OBS|IS,\theta_{s})$. The simplest assumption
is that if an irregularity exists, it will be detected, i.e., $P(OBS|IS,\theta_{s})\approx1$.
This assumption is justifiable for sufficiently low HF frequencies
with strong refraction in the ionosphere and for antennas with a wide
beamwidth in elevation plane. In such cases, the condition of orthogonality
to the magnetic field (which often necessary for effective scattering
from the most commonly observed irregularities elongated with 
the Earth magnetic field) will be satisfied in some part of the propagation
trajectory. These conditions are satisfied well for coherent scatter HF
radars of SuperDARN and SECIRA networks used in the paper.

Thus, the model $P(IS|\theta_{s})$ can be represented as the product
of the probability of observing signals by radars, $P(OBS|\theta_{s})$,
and the probability that the observed signal is one
scattered by the ionosphere, $P(IS|OBS,\theta_{s})$:

\begin{equation}
P(IS|\theta_{s})\approx P(IS|OBS,\theta_{s})\cdot P(OBS|\theta_{s})\label{eq:BayesEqSimplified}
\end{equation}

The probability of observing signals by radar network, $P(OBS|\theta_{s})$,
taking into account that SuperDARN and SECIRA radars have close characteristics,
is an experimentally measurable function. The probability
$P(IS|OBS,\theta_{s})$ depends on the specific way we detect signals
scattered by the ionosphere in signals observed by the radar. Thus,
we need to construct two models: the first (for $P(OBS|\theta_{s})$)
predicts where, when, and under what conditions we observe signals
by radars, while the second (for $P(IS|OBS,\theta_{s})$) allows us
to determine what fraction of the received signals is ionospheric
scattering.

\subsection{Detection of ionospheric scattering}

The problem of detecting ionospheric scattering in HF is quite complex,
and numerous algorithms currently exist for separating signals into
IS and non-IS components at SuperDARN and SECIRA radars. We tested several of them to
analyze common features and differences in the resulting probabilities
$P(IS|OBS,\theta_{s})$. To get statistics of ionospheric scattering
observations and make a preliminary analysis, we evaluated
three algorithms, including both standard and qualitative approaches.

To geolocate ionospheric irregularities, it is necessary to know radiowave
propagation trajectory. Therefore, accurate geolocation is performed
for signals with a known angle of arrival. The raytracing algorithm 
is the same as \cite{BERNGARDT_2026} takes
into account the actual ionosphere using ionospheric model IRI-2020\cite{IRI2020},
and does not account for the magneto-ionic effects. The elevation
angle calibration method required for these trajectory calculations
is described in detail in \cite{BERNGARDT_2026} and is based on an
analysis of meteor trail scattering heights. Based on the ray-tracing results,
we determined the geographic coordinates of the ionospheric scatterers,
and convert them to geomagnetic
coordinates using the AACGMv2 model\cite{AACGMv2}. So our dataset used for study 
is limited by the signals successfully processed by the FITACF algorithm
and successfully raytraced.

The classical widely used algorithm for IS detection (later in the paper - FITACF algorithm, FA) \cite{Blanchard_2011},
requires sufficiently high values of the Doppler velocity |V|
or spectral width |W| of the signal in velocity
units, to identify signal as IS:

\begin{equation}
|V|+A|W|>B;A=1/3;B=30m/s\label{eq:FAeq}
\end{equation}

The values of the parameters A and B in formula (\ref{eq:FAeq}) may
vary from one implementation to another; the values in (\ref{eq:FAeq})
are taken from the standard FITACF code.

The next algorithm (Convection Algorithm, CA) is heuristic one and does not
utilize spectral width information. Instead, it classifies any signal
with a Doppler shift exceeding 100 m/s ($|V|>100m/s$) as ionospheric
scatter. Such a high velocities are most common in high-latitude ionospheric
convection regions and almost never observed in other signals. So
the method should reproduce well-known statistical patterns associated with
the dynamics of the intensive high-latitude ionospheric convection.
Consequently, mid-latitude iononospheric scatter at MLAT<55, which is
detected by the FA algorithm, will be detected rarely.

The next algorithm is the 36-class classification (36C) uses self-learned neural
network to predict the scattering class \cite{BERNGARDT_2026}. The
three classes, previously interpreted as ground scatter (classes \#5,
12, and 15) are intepreted as not-IS signals while the remaining classes
are attributed to IS. In this case, the statistics of ionospheric
irregularities show smaller probability at high latitudes than FA. 
The overestimating IS at midlatitudes and underestimating
IS at high latitudes in comparison with FA may be due to the algorithm
\cite{BERNGARDT_2026} significantly underestimating the number of
signals scattered from the Earth's surface, and some other signals types 
could be non-IS signals.

To create dataset for analysis, for all the data listed in Table \ref{tab:RadarsYears}
we get the signals significantly exceeding the noise level (> 3 dB) and determine
the hourly average probability of observing ionospheric echoes $P(IS|OBS,MLAT,MLT)$
on a 1-degree geomagnetic latitude grid for each hour of observation.
The geographic scattering location was obtained from raytracing the signals using ionospheric model IRI-2020\cite{IRI2020}
and transformed to geomagnetic coordinates (MLAT, MLT) using AACGMv2 model\cite{AACGMv2} (see Fig.\ref{fig:RadarMap}B). 
The data used in the paper 
contains only the data with raytraced altitudes of less than 500 km. 
Only the data with a radar
range exceeding 450 km were used to exclude the signals scattered in the
vertical lobe of the antenna pattern, as well as those from sporadic
E layers and from meteor trails. The final dataset contains
approximately 12.5 million records.

\begin{figure}
\includegraphics[scale=0.7]{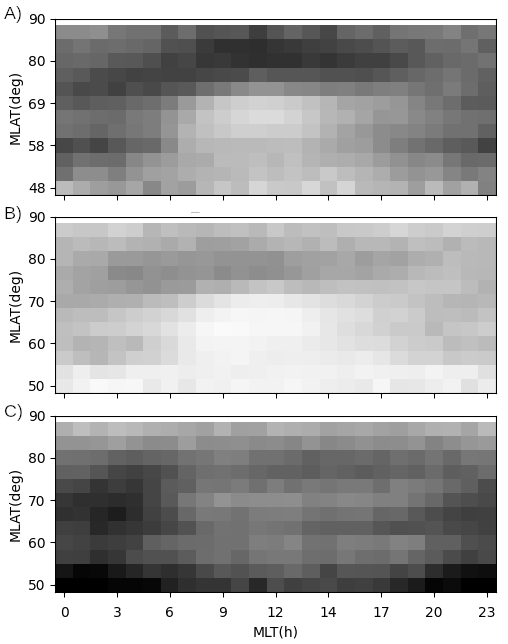}

\caption{Statistics of observations in MLAT,MLT
of ionospheric scattering according to three ionospheric
echo detection algorithms: (A) - FA; (B) - CA; (C) - 36C.}
\label{fig:IonoStatsAlgos}
\end{figure}

Statistics of ionospheric echo probabilities $P(IS|OBS,MLAT,MLT)$, obtained by applying
the three algorithms described above to the entire dataset, are presented in geomagnetic coordinates
in Fig.\ref{fig:IonoStatsAlgos}.

Figure \ref{fig:IonoStatsAlgos} shows that algorithm results for
detecting IS can be described as follows.
The algorithm 36C pays significant
attention on radio wave propagation characteristics and minimizes
the number of signals interpreted as non-IS (i.e.ground scatter),
thereby showing a high probability that an observed signal is an
ionospheric echo. The algorithm CA pays significant
attention to high drift velocities and minimizes the number of signals
interpreted as ionospheric scatter. It demonstrates the known dynamics
of the equatorial boundary of convection zone (Heppner-Maynard), reaching
its highest latitude around 12 MLT and it shows low probability
that an observed signal is an ionospheric echo.
The standard FA algorithm is nearly between these two algorithms, demonstrating
intermediate properties - specifically, a moderate degree of polar
oval dynamics and a moderate probability that the observed signal
is an ionospheric echo. Therefore, it is worthwhile to analyze FA
detecting technique, as it produce some intermediate results.

\subsection{Model input parameters $\theta_{s}$}

Indices of solar and geomagnetic activity, as well as solar wind satellite data, 
were used as input parameters $\theta$ for the model.
The well known dependence of the polar oval on geomagnetic coordinates
allows us to include magnetic local time and magnetic latitude as 
the model's input parameters, together
with geographic latitude. All these parameters together forms the
parameters vector $\theta_{s}$ for the model. 
Hourly averaged space weather parameters are used as model inputs for forecasting
the probability of observing ionospheric irregularities $P(IS|OBS,\theta_{s})$.
The parameters used for the following analysis in the paper are listed
in Table \ref{tab:InputIndexes}.

\begin{table}
\caption{Parameters used for modeling}

\begin{tabular}{|p{2cm}|p{3cm}|p{7cm}|}
\hline 
Parameters & Comments & Data source\tabularnewline
\hline 
\hline 
\centering
$K_{p}$, $a_{p}$ &
\centering
Geomagnetic disturbance indices &
\centering
\cite{HelmholtzCentreForGeosciences}
\tabularnewline
\hline 
\centering
F10.7 & 
\centering
Solar activity index & 
\centering
\cite{HelmholtzCentreForGeosciences}
\tabularnewline
\hline 

\centering
AE, AU, AL, AO
 & 
\centering
Auroral geomagnetic disturbance indices
 & \cite{KyotoWDC}\tabularnewline
\hline 

\centering
ASY-H, ASY-D,
SYM-D, SYM-H
 & 
\centering
Mid-latitude geomagnetic 
disturbance indices
 & \cite{KyotoWDC}\tabularnewline
\hline 

\centering
$B_{x}$, $\widetilde{B_{y}}$, $\widetilde{B_{z}}$,
$B_{t}$, $\psi_{CA}$
 & 
\centering
Interplanetary magnetic 
field parameters in 
GSM coordinates and
IMF clock angle
 & \cite{OMNIdatabase}\tabularnewline
\hline 
\centering
$V_{x}$, $V_{y}$, $V_{z}$ & 
\centering
Solar wind speed 
in the longitudinal (x) 
and transverse (y, z) 
directions.
 & \cite{OMNIdatabase}\tabularnewline
\hline 

\centering
$\rho_{sw}$, $T_{sw}$,
$P_{sw}$, $E_{sw}$
 & 
\centering
Solar wind parameters: 
density, temperature, 
pressure, and 
electric field amplitude.
 & \cite{OMNIdatabase}\tabularnewline
\hline 

\centering
MLT, MLAT,
GLAT
 & 
\centering
Geomagnetic and geographic 
coordinates of the 
scattering point
 & 
\centering
Calibrated radar data 
from Table \ref{tab:RadarsYears},
IRI-2020 and AACGMv2 models.
\tabularnewline
\hline 
\end{tabular}\label{tab:InputIndexes}
\end{table}

We used the following additional notations:
\begin{itemize}
\item IMF clock angle: $\psi_{CA}=atan(B_{y}/B_{z})$;
\item Tangential component of the IMF: $B_{t}=\sqrt{B_{y}^{2}+B_{z}^{2}}$;
\item Normalized projection of the IMF onto the y-axis: $\widetilde{B_{y}}=B_{y}/B_{t}=sin(\psi_{CA})$;
\item Normalized projection of the IMF onto the z-axis: $\widetilde{B_{z}}=B_{z}/B_{t}=cos(\psi_{CA})$;
\item Electric field amplitude: $E_{sw}=-\left|V\right|B_{z}$.
\end{itemize}

Thus, the model can be constructed using 23 input parameters describing
space weather conditions on 1-hour time scales, and three spatial location
parameters - MLAT, MLT, and GLAT.

\subsection{Optimal coordinate system}

The preliminary step of analysis is to find an optimal coordinate system
for $P(IS|OBS,\theta_s)$ prediction. We consider two variants: related with geographic
latitude GLAT and with geomagnetic latitude MLAT. 
To select the optimal coordinate system, we trained a simple
non-linear regression model (a four-layer fully connected neural network
with 3 hidden layer of 64 neurons and absolute activation function) that takes the parameters from
Table \ref{tab:InputIndexes} as input and predicts $P(IS|OBS,\theta_{s})$.
This approach relies on the assumption that all the space weather parameters cause irregularity
generation without any time-lagged effects.

When building and training the simple neural network we used 
the following useful approaches, which we also used when building and
training much more complex final model.

\textbf{Input data transformation (quantile transformation).} Most
measured physical parameters are characterized by two features: they
are measured with error and they may have a wide dynamic range
(up to several orders of magnitude). To build robust model, we include
into the model input data transformation refered as quantile transformation (QT).

In the first stage, the transformation interpolates 'bad'
or missing data in each space weather parameter with the parameter's previous
'good' value. This ensures data continuity
and fills data gaps - a critical step for future constructing a comprehensive
model that accounts for observational history. This approach is also
critical during extreme disturbances when certain space weather parameters cannot
be correctly measured and public databases contains
special 'error' values. We hope that rare errors caused by this interpolation
is compensated by the large volume of data used for the analysis.

In the second stage we transform data from wide dynamic range to small limited range.
To do this the positions of quantile positions within the
data values were determined for each input space weather parameters 
and coordinates MLAT,MLT,GLAT: a maximum of 40 quantiles for latitudes
and time, and a maximum of 25 quantiles for space weather data. 
In the cases where the necessary number of qunatiles could
not be found (for example, for discrete variables with fewer number
of values, such as $K_{p}$ or hours of MLT), the actual number of
descreet values for that variable was used.
In this stage, the output data were linearly interpolated between
the quantile positions to ensure the continuity of the input data
transformation. As a result we constructed monotonically increasing, 
piecewise-linear function $f_{i}$ for each i-th
input parameter $\theta_{s,i}$. This function transforms each input parameter 
to output value $\theta^{f}_s$ in the range of $[0,40]$ (or a narrower range) to
the distribution density of $\theta^{f}_s$ across the dataset to be
approximately uniform. 
This transformation limits the variation range of
the following network's input parameters $\theta^{f}_s$ and therefore enhances the algorithm's robustness
when handling future unknown data.

The stage can be described by the following:

\begin{equation}
\begin{array}{l}
\theta^{f}_{s,i}(t)=\left\{ \begin{array}{l}
\theta_{s,i}(t),if\,\theta_{s,i}(t)\,is\,correct\,value\\
\theta^{f}_{s,i}(t-\Delta t),\,in\,other\,cases
\end{array}\right.\\
QT_{s}(\theta{}_{s,i})=f_{i}(\theta^{f}_{s,i})\\
\end{array}
\label{eq:QT_all}
\end{equation}

where $f_{i}(x)$ is a monotonically increasing function 
constructed using the algorithm described above:

\begin{equation}
f_{i}(\theta_i)=j+\frac{\theta_i-Q_{i,j}}{Q_{i,j+1}-Q_{i,j}};if\,\theta_i\in[Q_{i,j},Q_{i,j+1})
\label{eq:f_q}
\end{equation}

and $Q_{i,j}$ is a j-th quantile position for i-th input parameter $\theta_{i}$,
constructed using quantile discretization:

\begin{equation}
\begin{array}{l}
P(\theta_{i}\in(Q_{i,j},Q_{i,j+1}])\approx P(\theta_{i}\in(Q_{i,k},Q_{i,k+1}])\\
\forall j,k\in[0,N_{i})\\
N_{i}\leq25
\end{array}
\label{eq:Q_QT}
\end{equation}

where $P(x \in I )$ is the probability that x belongs to the interval I.

\textbf{Data augmentation.} To enhance the model robustness and physical abequacy, a random
augmentation of the input data was performed.
As input $\theta^{a}$ of the neural network we use the augmented QT values:

\begin{equation}
\theta^{a}_{s,i}=QT_{i}(\theta_{s,i})+\varepsilon
\label{eq:augmentQT}
\end{equation}

where $\varepsilon$ is a uniformly distributed random variable in
the range {[}-1, 1{]}.

The transformed and augmented dataset obtained in this way is used
to train the model.

Using each of the three specified ionospheric scatter detection methods, the
neural network model described above was trained, and the significance
of each parameter was evaluated for each IS detection method. A permutation-based
method \cite{PermutationFI} was applied to the trained model to estimate significance
of each input parameter (feature importance, FI). In terms of permuation
method, the FI value is the value of decreasing model prediction quality
(we use mean squared error, MSE).

To obtain confidence intervals of feature importance, each model was tested 10 times, each
time using a randomly selected 20\% subset of the training dataset. 
The choice 
of the most important parameters was based on these test runs. 
This allows us to estimate that a parameter
is statistically significant at the 95\% confidence level (using approximate
condition $<FI>-2\sigma(FI)>0$, where $<x>$ - mean value of x, $\sigma(x)$ - its standard deviation over test runs).
Then we order the paramters by their significance $<FI>$.
Once a parameter in the ranked list
was found as insignificant, all the parameters with lower significance
were also considered insignificant.

The most important parameters for nowcast using simple neural nowcast network 
with a 1-hour averaging time are listed in Table
\ref{tab:HourlyImportanceGLAT}.

\begin{table}
\caption{Hourly-averaged parameters most significant for nowcasting
(in descending order of significance), using various
algorithms of detecting ionospheric scattering from mixture with other
signals (simple nowcast network).}

\begin{tabular}{|c|c|}
\hline 

\centering
IS detecting
algorithm
 & 
\centering
Significant parameters
\tabularnewline
\hline 
\hline 
FA & MLAT,MLT,AE,AL,$\widetilde{B_{z}}$\tabularnewline
\hline 
CA & MLAT,MLT,AL,GLAT,$\widetilde{B_{z}}$,AE,$E_{sw}$,$B_{t}$\tabularnewline
\hline 
36C & MLAT,GLAT\tabularnewline
\hline 
\end{tabular}\label{tab:HourlyImportanceGLAT}
\end{table}

The results presented in Table \ref{tab:HourlyImportanceGLAT} demonstrate
that, almost regardless of the ionospheric scattering detection method,
selecting the geomagnetic latitude (MLAT) as a model input parameter
is preferable to using the geographic latitude (GLAT). This is corresponds well 
with widely used approaches to the analysis of scattering regions in
the high-latitude ionosphere \cite{Ruohoniemi_1996}.

\section{Model for $P(IS|OBS,\theta_{s})$}

\subsection{Model architecture}

As previously noted, the probability of irregularity formation depends
on each space weather parameter dynamics in a different manner - 
this can be implemented by moving
average of the values \cite{McGranaghan_2018} or by finding
specific patterns \cite{Gao_2025}. Let us construct a model capable
of both generating the spatial distribution of irregularities and
identifying optimal temporal patterns within each input
parameter that improve forecast quality.

Forecasting the spatiotemporal distribution of irregularities has
significant problems: radar observations may be missing during certain periods
(e.g., due to radar downtime, signal absorption in the ionospheric
D/E layers or high noise level, radiowave propagation conditions or data unavailability);
radar fields of view do not cover the entire globe; and scattering
is observed in only small regions in radar field-of-view. Furthermore,
the dataset is partially ordered: while geophysical indices are fully
ordered in time, the sequence in which the radar observations are
sampled across Magnetic Latitude (MLAT) and Magnetic Local Time (MLT)
at any given moment can have arbitrary order. As a
result, we are dealing with a huge region containing irregularities
and attempting to reconstruct its shape and properties through a number
of small 'holes', whose number and location
change unpredicteably over the time. We never observe the whole region
in the same time. 

Developing a neural network capable to learn at such a dataset is
difficult. Traditional layered neural networks designed for forecasting
ordered data are unsuitable here. Recurrent networks, typically used
for sequence prediction, are inapplicable too, as the number of radar
data points and the order of its acquisition can vary arbitrarily
at any given moment. Modern generative diffusion networks like \cite{WeatherNext3}
are also unsuitable due to the extreme sparsity of the input data
- we are observing discrete points instead of image fragments.

\begin{figure}
\includegraphics[scale=0.2]{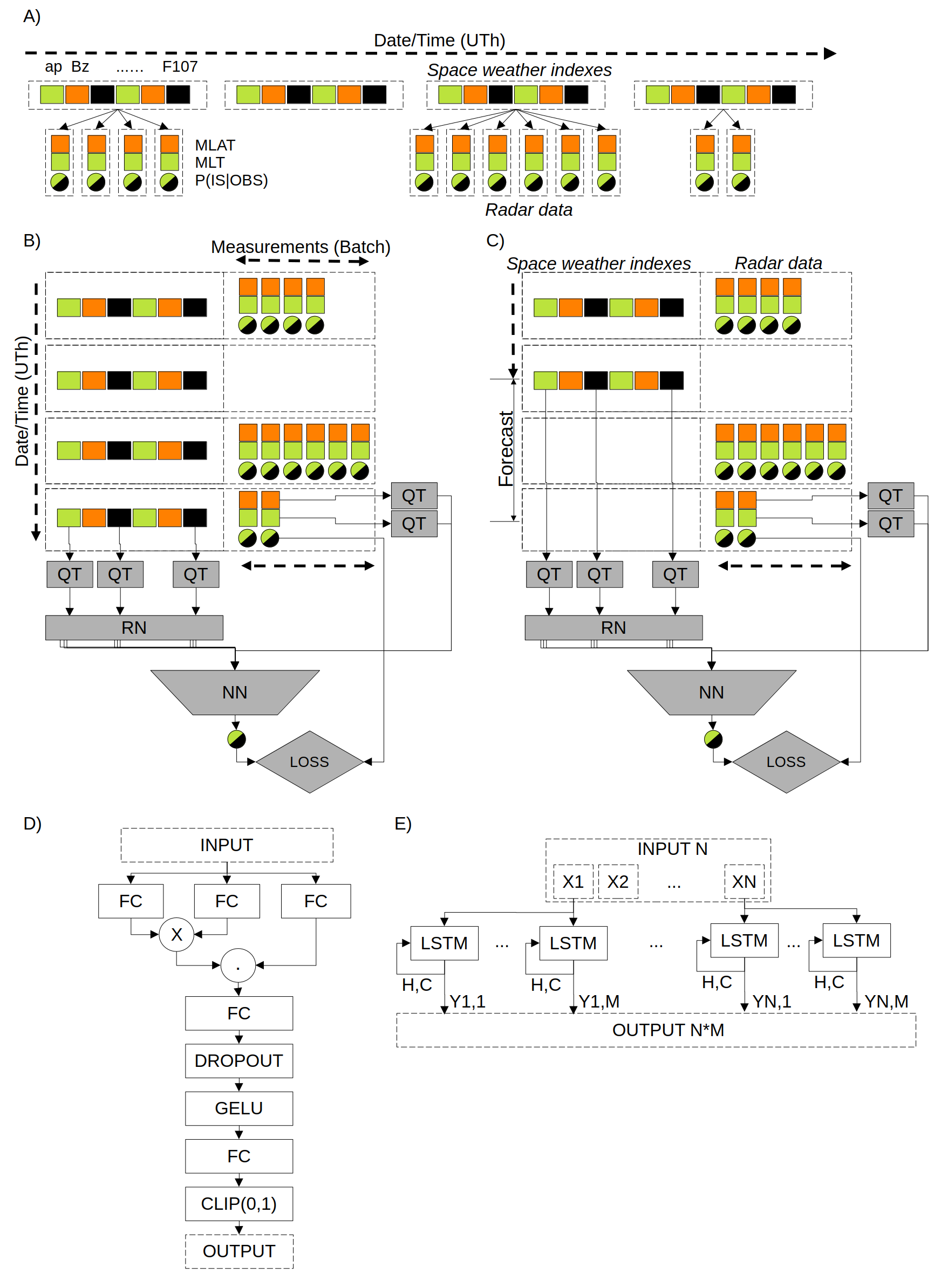}
\caption{(A) - dataset structure in the form of dynamic graphs, dashed lines
indicates nodes, arrows - directed edges; (B-D) - regression neural
network (RN-NN) for predicting the probability of ionospheric scattering
and its training method in nowcasting mode (B) and in forecasting
mode (C). Dashed lines indicate blocks fed into the network simultaneously.
Notation: QT - quantile transformer, RN - recurrent neural network,
NN - attention neural network. (D) - NN architecture. FC - standard
dense (i.e.fully connected) layer, GeLU - nonlinear activation function,
Dropout - standard dropout layer with dropout probability 0.3, Clip
- clipping values to the interval {[}0,1{]}. (E) - RN architecture.}
\label{fig:RNNN}
\end{figure}

In fact, such a dataset can be represented as a sequence of
graphs (or dynamical graphs), as shown, for example, in \cite{Gao_2025}.
Therefore using graph neural networks \cite{GNN} looks a good choice for the
task. However, our graphs actually have a very simple and similar
structure (shown in Fig.\ref{fig:RNNN}A): a level-1 tree where
the root node holds geomagnetic and solar parameters, while the numerous
branches contain only the radar data - geomagnetic coordinates and
the probabilities $P(IS|OBS,\theta_s)$ we try to predict. Consequently,
the graph neural network architecture can be significantly simplified.

To build the model, an architecture was created reffered to the regression-neural
network (RN-NN). It represents an enhanced and simplified version
of a recurrent graph attention neural network. The model can be trained
on such a dataset taking into account its 'graph' characteristics:
the ordering of graphs over the time variables, and unordered structure
over spatial variables. The model architecture is shown in Fig.\ref{fig:RNNN}B.

In the paper we will construct and train models for both nowcasting
and forecasting. We define nowcasting model as a model that, based
on input values measured up to a specific moment, predicts the distribution
of ionospheric irregularities over the entire hemisphere (the
mid- and high-latitude ionosphere of the Northern Hemisphere) at that
moment. The nowcasting network is trained according to the scheme
shown in Fig.\ref{fig:RNNN}B.

We define forecasting model with a lead time T hours as a model
that, based on input values measured up to a specific moment, predicts
the distribution of ionospheric irregularities across the entire hemisphere 
for the time T hours after that given moment. Nowcasting
differs from forecasting only in the latest record of the input parameters 
fed into the RN part of the network. The greater the 
difference between the date/time for the RN component and the date/time
for NN component, the longer the forecast lead time. The forecast network
has the same architecture as nowcast network, but
is trained using the the scheme shown in Fig.\ref{fig:RNNN}C.

The figure shows that both architectures (the nowcasting model and
the forecasting model) are identical, which is typical for graph neural
networks - their architecture generally
does not depend on the number of nodes in the graph or the forecast
lead time. Let us discuss the network architecture and training details.

\textbf{Model training notes.} When working with graph data using
graph neural networks, the concept of nodes neighborhood is crucial:
the predicted state of each node is determined by its own parameters
and those of all its neighbors. Each data graph at a given time step
is directed. So for every radar data node (a point in MLAT,MLT space),
where we try to predict the $P(IS|OBS,\theta_s)$ there is only single neighbor
node: the space weather parameters node. Within the framework of the powerful and widely used
GATv2 architecture \cite{GATv2}, for each node we must combine (concatenate) the
properties of that node (MLAT, MLT) with those of its neighbor (the
space weather parameters), then perform an attention operation \cite{Attentionneed2023}, and
then apply a small decision neural network to the result. This operation must
be carried out simultaneously for all nodes (all the radar data points).
The result is a graph containing predictions of the target parameter
value for each node $P_{pred}(IS|OBS,\theta_{s})$. This prediction can be
compared with the experimentally observed $P_{obs}(IS|OBS,\theta_{s})$ value.
Minimization of a residual functional allows training the whole neural network. 
The network architecture is shown in Fig.\ref{fig:RNNN}B.
It should be noted that when processing graph data, all nodes in the
graph are processed simultaneously. So the result is independent on the processing order.
In standard machine
learning terms, all nodes of a single graph forms a processing batch
that is invariant to the order of its samples. This is illustrated
in Fig.\ref{fig:RNNN}B-C by arrows, showing training order. 

The next feature of the model is the prediction of node properties
for the current (nowcast mode) or subsequent (forecast mode) graph based on a sequence of graphs
(time steps; see Fig.\ref{fig:RNNN}A). Since we are dealing with
dynamic graphs, it is necessary to propagate the history of node states.
Obviously, the only single node consistently present across all our
data graphs - it is the node containing space weather parameters.
Therefore, we pass its state history through a recurrent neural network
based on an LSTM unit. This network increases the number of features
of the primary node, it is a standard technique in graph neural networks
that enhances their quality. The architecture of this network is illustrated
in Fig.\ref{fig:RNNN}E.

\textbf{Quantile transformers (QT)} were previously described 
in (\ref{eq:QT_all}-\ref{eq:Q_QT}).
Thus, the input fed to the RN and NN networks at any given moment is not the
raw measured data itself, but rather a limited monotonic function
of that data without error values and with nearly uniform distribution.

Each \textbf{regression network (RN)} has a 1-to-1 recurrent neural
network (RNN) architecture, where the value of the final element in
the output sequence is selected as the output. In the RNN an LSTM
unit chosen for its robustness in training at long sequences (the
history sequence length used is 31 days, or 744 values). A separate
multi-head RNN is implemented for each input parameter with six heads selected
experimentally to maximize forecast quality. 
Consequently, at any given time moment, we get six output values for every input parameter. 
These
can be interpreted as the result of filtering of that
parameter's value sequence by six independent non-linear filters. During
training, these filters adapt to identify signifficant patterns in
the parameter to provide the optimal forecast quality by the resulting
neural network. The architecture (Fig.\ref{fig:RNNN}E) was selected to maximize the
isolation of each parameter and its filters. Later this allows making effective
future analysis of most important parameters and enhancing the model's
interpretability.

A drawback of RN model (Fig.\ref{fig:RNNN}E) is its slow
inference and training speed within the standard Python/PyTorch framework 
that makes it slow to train and study. 
However, the model has obvious symmetry leading to computational parallelism,
which allows optimization when running on GPUs. Therefore the RN network code was
rewritten and parallelized from Python/PyTorch framework to Python/C++/CUDA(r)
code using recent state-of-the-art large language models (Opus5 and Qwen3.8Max). 
The new code shows a tenfold increase of neural network
training speed and allows to complete all training and analysis of the network 
described in the paper using regular personal computer with regular GPU.

\textbf{The decision-making neural network (NN)} takes as input parameters
the results calculated by regression network (RN), six output features
per each input space weather parameter (derived from six parameter filtering
operations) and uses them for each of radar observation point (MLAT,
MLT) to predict the probability of observing ionospheric scattering
$P(IS|OBS,\theta,MLAT,MLT)$ at spatial location MLAT,MLT.

As already discussed, the NN operates on unordered radar data (scatering
positions) for given moment. In contrast, the RN operates on the ordered
subset of the dataset (space weather parameters), processing it
in strict temporal order, and this allows the extraction of temporal
patterns that are most significant for the prediction.

The neural network architecture employs a classic attention mechanism \cite{Attentionneed2023} 
followed by two-layer fully connected network with dropout,
and is shown in Fig.\ref{fig:RNNN}D.

From machine learning point of view
the suggested network Fig.\ref{fig:RNNN}B-E is an implementation 
of recurrent graph attention network for predicting
node parameters over dynamic graph sequence. 
From physical point of view the network is a model that
predicts ionospheric scattering probability using  
scattering point geomagnetic coordinates and
several non-linear filterings of each space weather parameter 
temporal sequence.

\subsection{Model training and simplification}

For training the model, the data were split into two parts: a training
dataset and a validation (test) dataset, as shown
in Table \ref{tab:RadarsYears}. The validation dataset (2024-2025)
followed the training dataset (2010-2023) chronologically. 

\textbf{Compensating for data incompleteness during training.} A problem
of the model training is the incompleteness of the radar data: the radar networks
do not provide a hemisphere coverage. This makes impossible to determine correctly 
the ionospheric irregularities at low-latitudes. 
After model training, this results in cases where
irregularities are predicted at all latitudes down to equatorial,
which cannot be verified using existing radar experimental data. 

To compensate this problem, the two preprocessing steps was introduced
prior to data augmentation. At the first step for the lowest 2.5\%
of magnetic latitudes in the dataset, the observed occurrence rate of irregularities
was forced to zero while keeping all other parameters identical to
the experimental observations. At the second step, we extend dataset with zero occurence rate samples
in the mid-latitude band. Namely an identical dataset of size 10\%
was randomly choosen from training dataset, but with magnetic latitude
and ionospheric scattering probability are modified: the latitude was replaced
with a random value within the lowest 10\% of the latitude range (approximately
4 degrees above the lowest latitude boundary), and $P(IS|OBS,\theta_{s})$ was set to zero. 
This approach
stimulates models to decrease predicted $P(IS|OBS,\theta_{s})$ at
lowest measured latitudes.

Both types of networks (RN and NN)
are trained jointly and simultaneously. Consequently, the search for
temporal patterns in the space weather parameters and the method for estimating $P(IS|OBS,\theta_s)$ 
based on these patterns are optimized together. 

The training algorithm is as follows.
\begin{enumerate}
\item During the training epoch, we iterate sequentially through all moments in the train dataset
from the earliest to the latest.
\item Select next time moment T.
\item If no radar data is available for this moment, go to step 2.
\item Pass the input parameters throught QT and augment the output. To accelerate computations,
we use a one-month window $(T-t)\leq31\times24$ h.
\item Pass the result to the RN and compute the output based on the history
of parameters $\theta$ ordered by time $t\le T$. Store the resulting RN output values $\phi$
Note that these values are part of the computation
graph, so the RN will be optimized during training at step 9.
\item Pass the features (MLAT, MLT) through QT for each existing radar observation node for given moment, 
augment the results to compensate data incompleteness.
\item Concatenate the results with the output features $\phi$ of
the regression network RN (computed in the step 5)
and form a batch of inputs $\chi$ for the neural network NN;
\item Pass the batch of results through NN.
\item Calculate the difference between the NN outputs $P_{pred}(IS|OBS,\theta,MLT,MLAT)$
and the experimental observations $P_{obs}(IS|OBS,\theta,MLT,MLAT)$
using the mean squared error (MSE) loss function, and perform a training
step on the resulting batch. Simultaneously update all coefficients
of the regression network (RN) and the neural network (NN), using 
the AdamW optimization algorithm with a learning rate of $10^{-5}$.
\item If next time moment exists go to step 2.
\item Calculate loss at whole validation dataset. 
If the loss on the validation dataset is minimized, exit the training
loop, otherwise go to step 1 and train next epoch.
\end{enumerate}
Thus, training regression network (RN) proceeds sequentially, record
by record of space weather parameters, whereas the neural network
(NN) trains batch-wise using radar data; the batch size varies over
time, corresponding to the number of simultaneous scattering observations
across the radar network at the given moment. In all the cases, augmentation
of space weather parameters and coordinates is performed by no more than +/-1.0 of the output discrete
values (\ref{eq:augmentQT})after quantile transformer QT.

Training typically converges in fewer than 50 epochs, due to the model's
simplicity, the large size of the dataset, and the
high level of data augmentation.

\begin{figure}
\includegraphics[scale=0.25]{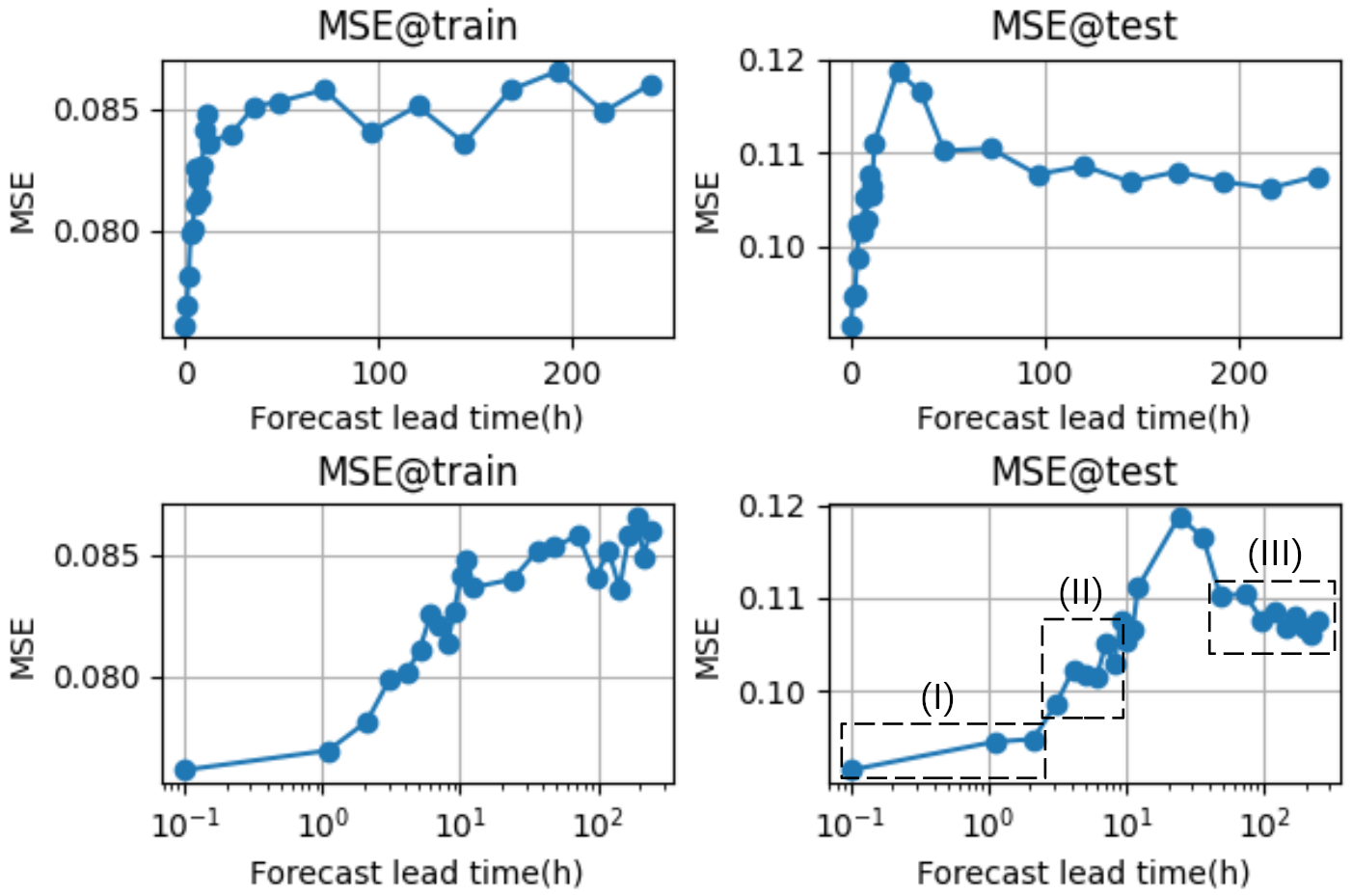}

\caption{Comparison of forecast model quality at different lead times. 
Marked regions are: (I) - nowcast and short-term forecast; (II) - mid-term forecast; (III) - long-term forecast.}
\label{fig:fit_Q}
\end{figure}

A separate models were trained for every forecast lead time
to enable subsequent analysis of the space weather indices most important
for the irregularities forecast. Figure \ref{fig:fit_Q} shows the
model's mean squared error (MSE) for lead times ranging from 0 (nowcasting)
to 10 days (240 hours) at train and validation (test) datasets.

The Fig.\ref{fig:fit_Q} shows three lead intervals of acceptable forecast performance:
nowcasting and short-term forecasting (0-2 hours), mid-term
forecasting (3-9 hours), and long-term forecasting (4-10 days).
The 10-72 hour lead-time range is characterized by the highest error
rates and due to poor predictability it should not be used.

The Fig.\ref{fig:fit_Q}D shows that nowcasting demonstrates the highest forecast quality,
with comparable performance at lead times of up to two hours. Forecast
error increases as the lead time increases.

\begin{figure}
\includegraphics[scale=0.5]{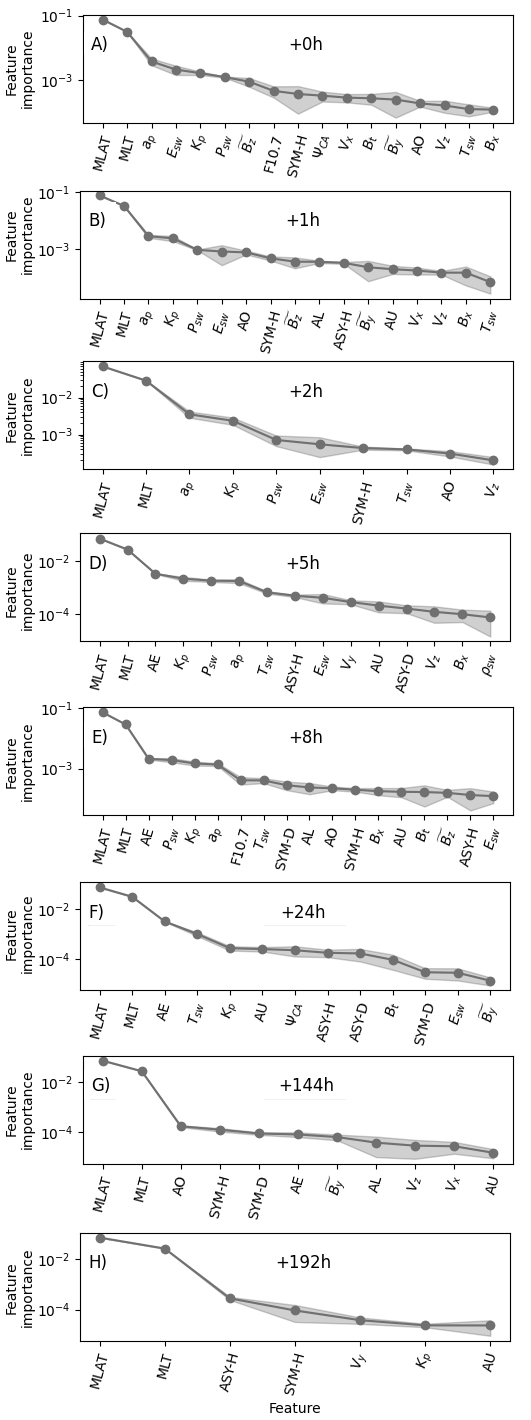}
\caption{Importance of model parameters for different forecast lead times (from
0 hours to 192 hours)}
\label{fig:fi_forecast}
\end{figure}

To find the most important parameters for prediction we use permutation technique discussed before.
Fig.\ref{fig:fi_forecast} illustrates the relative importance
of various model input parameters (components of $\theta_{s}$) for
different lead times. Table \ref{tab:fi_forecast_best} lists the
most significant parameters. The table shows that the most important parameters are always MLAT,MLT.

\begin{table}
\caption{Most important model parameters by forecast lead time (MLAT and MLT
excluded), in descending order of importance}

\begin{tabular}{|c|c|}
\hline 
Forecast lead time(h) & Important parameters\tabularnewline
\hline 
\hline 
0 & $a_{p},E_{sw},K_{p},P_{sw},\widetilde{B_{z}}$,F10.7\tabularnewline
\hline 
1 & $a_{p},K_{p},P_{sw},E_{sw}$,AO\tabularnewline
\hline 
2 & $a_{p},K_{p},P_{sw},E_{sw}$,SYM-H\tabularnewline
\hline 
3 & $a_{p},K_{p},T_{sw},P_{sw}$,AE\tabularnewline
\hline 
4 & $a_{p},K_{p}$,AE,$P_{sw}$,ASY-H\tabularnewline
\hline 
5 & AE,$K_{p},P_{sw},a_{p},T_{sw}$\tabularnewline
\hline 
6 & $T_{sw},K_{p},a_{p},P_{sw}$,AE\tabularnewline
\hline 
7 & AE,$K_{p},a_{p}$,ASY-H,$T_{sw}$\tabularnewline
\hline 
8 & AE,$P_{sw},K_{p},a_{p}$,F107\tabularnewline
\hline 
9 & AE,$a_{p},P_{sw},K_{p},V_{y}$\tabularnewline
\hline 
12 & $P_{sw}$,AU,$K_{p}$,AL,$a_{p}$\tabularnewline
\hline 
24 & AE,$T_{sw},K_{p}$,AU,$B_{tan}$\tabularnewline
\hline 
36 & AE,ASY-D,F107,$P_{sw}$,AU\tabularnewline
\hline 
144 & AO,SYM-H,SYM-D,AE\tabularnewline
\hline 
168 & ASY-D,$T_{sw}$\tabularnewline
\hline 
192 & ASY-H,SYM-H\tabularnewline
\hline 
\end{tabular}\label{tab:fi_forecast_best}
\end{table}

The table shows that the short-term forecast (up to 2 hours) is well
determined by the parameters $a_{p}$, $K_{p}$, $P_{sw}$, and $E_{sw}$,
while $\widetilde{B_{z}}$ and F10.7 are also necessary for nowcasting (0 h)
only. The mid-term forecast (3-6 hours) is determined by the
parameters $a_{p}$, $K_{p}$, $T_{sw}$, $P_{sw}$, and $AE$. For
the 7-9-hour forecast, the prediction is based on the parameters
$AE$, $a_{p}$, $K_{p}$, and $P_{sw}$. For longer-term forecasts (above 9 hours),
no specific parameter structure is observed. Thus, the parameters
$a_{p}$, $K_{p}$, $P_{sw}$, $E_{sw}$, $\widetilde{B_{z}}$, $T_{sw}$,
$AE$, and $F10.7$ can be selected as the base parameters for models
providing forecasts with lead times from 0 to 9 hours.

\begin{figure}
\includegraphics[scale=0.9]{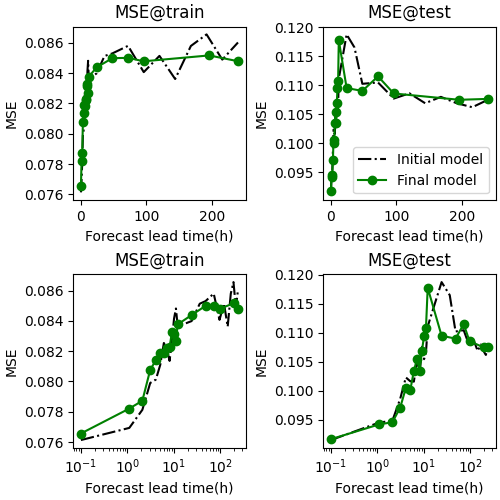}

\caption{Training quality of distilled models at train and test datasetfor
various lead times in comparison with initial model: initial model
- 23 input parameters, final distilled model - 8 parameters: $a_{p},K_{p},P_{sw},E_{sw},\widetilde{B_{z}},T_{sw}$,AE,F10.7}
\label{fig:distilated}
\end{figure}

After identifying the model's most important input parameters, the
number of them was reduced, and the models Fig.\ref{fig:RNNN} were trained using only
these (distilled) parameters on the same dataset for various forecast
lead times. A comparison of the quality of the final distilled models
versus the original ones is shown in Fig.\ref{fig:distilated}.

The figure shows that, within the 0-9 hour forecast horizon, the
model based on the parameters $a_{p},K_{p},P_{sw},E_{sw},\widetilde{B_{z}},T_{sw}$,
AE, F10.7 performs no worse than the original model on the validation dataset,
confirming the effectiveness of the selected parameters for forecasting.
Attempts to reduce the number of input parameters lowers the
model quality, demonstrating that this set of parameters looks minimal
for effective nowcasting and forecasting 
among all the studied space weather parameters (Table \ref{tab:InputIndexes}).

\section{Model for $P(OBS|\theta)$}

To construct a complete model of the occurrence of ionospheric irregularities
$P(IS|\theta_{s})$ one need not only $P(IS|OBS,\theta_{s})$ model,
built above, but also forecast of the probability that radars observing
various signals under different conditions, $P(OBS|\theta_{s})$.
The combination of these two models (\ref{eq:BayesEqSimplified}) 
allows one to estimate $P(IS|\theta_{s})$.

To construct the model $P(OBS|\theta_{s})$ based on the dataset previously
used to build the $P(IS|OBS,\theta_{s})$ model, a second dataset
was created. For each moment of observations (date and UT hour) 
in the training dataset,
a binary map of total received signals (IS and non-IS) observations was generated, 
irrespective of the specific signal type.
The data are presented as the grid $(MLAT,MLT)=([40,90],[0,24])$. 
The 0 at this map indicates the absence of any received
signal by radars at given (MLAT,MLT) , while 1 indicates the presence
of at least one signal at the given moment and location.
During followed training, binary signal occurrences on this grid becomes averaged, forcing
the model to predict the target probability $P(OBS|\theta_{s})$. 
As a model we use an architecture identical to the RN-NN 
in nowcasting mode (Fig.\ref{fig:RNNN}B), 
but train it on the dataset described above.

As shown the neural network training, the probability $P(OBS|\theta_{s})$
depends weakly on the space weather state $\theta$ and the local
magnetic time MLT, and can be represented in a much simpler
form:

\[
P(OBS|\theta_{s})\approx P(OBS|MLAT)
\]

\begin{figure}
\includegraphics[scale=1.0]{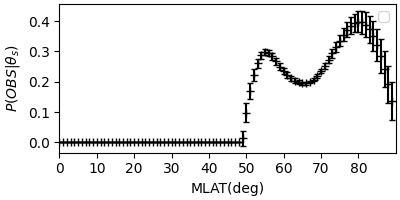}

\caption{Prediction statistics of $P(OBS|\theta_{s})$.
Intervals corresponds the mean results, error interval
shows standard deviation over the dataset.}

\label{fig:cmp_Pobs}
\end{figure}

To demonstrate this a distribution of the model results in different conditions 
shown in Fig.\ref{fig:cmp_Pobs}. The figure shows the mean
predicted probability and the standard deviation calculated over a
dataset containing roughly equal numbers of randomly selected profiles
for each $K_{p}$ level. This random selection is made to compensate
for potential errors arising from dataset non-uniformity: in the original
dataset low $K_{p}$ values occur several orders of magnitude more
frequently than high $K_{p}$ values. The figure demonstrates that
the models makes qualitatively consistent predictions, showing two
primary peaks in scattered signal observations at latitudes of 55
MLAT and 80 MLAT with a minimum between them, and
sets $P(OBS|MLAT)$ to zero at latitudes below 50 MLAT.

The figure shows that the RN-NN model demonstrates little variability
in response to changing external conditions $(\theta,MLT)$, with
the typical probability $P(OBS|MLAT)$ ranging from 0.2 to 0.45 over
the magnetic latitude range from 55 to 90 MLAT. Therefore, to describe
$P(OBS|MLAT)$ in the subsequent analysis, we use a stationary tabulated
profile of $P(OBS|MLAT)$ predicted by the RN-NN model and averaged
over the available dataset covering various space weather conditions
$\theta$ and MLT.

\section{Resulting model $P(IS|\theta_{s})$ and discussion}

The final model $P(IS|\theta_{s})$ is formed as the product of the
tabulated model for the observation probability of scattered signals
$P(OBS|MLAT)$ and the probability that a scattered signal is one
scattered from the ionosphere $P(IS|OBS,\theta_{s})$, following (\ref{eq:BayesEqSimplified}).

\begin{figure}
\includegraphics[scale=0.4]{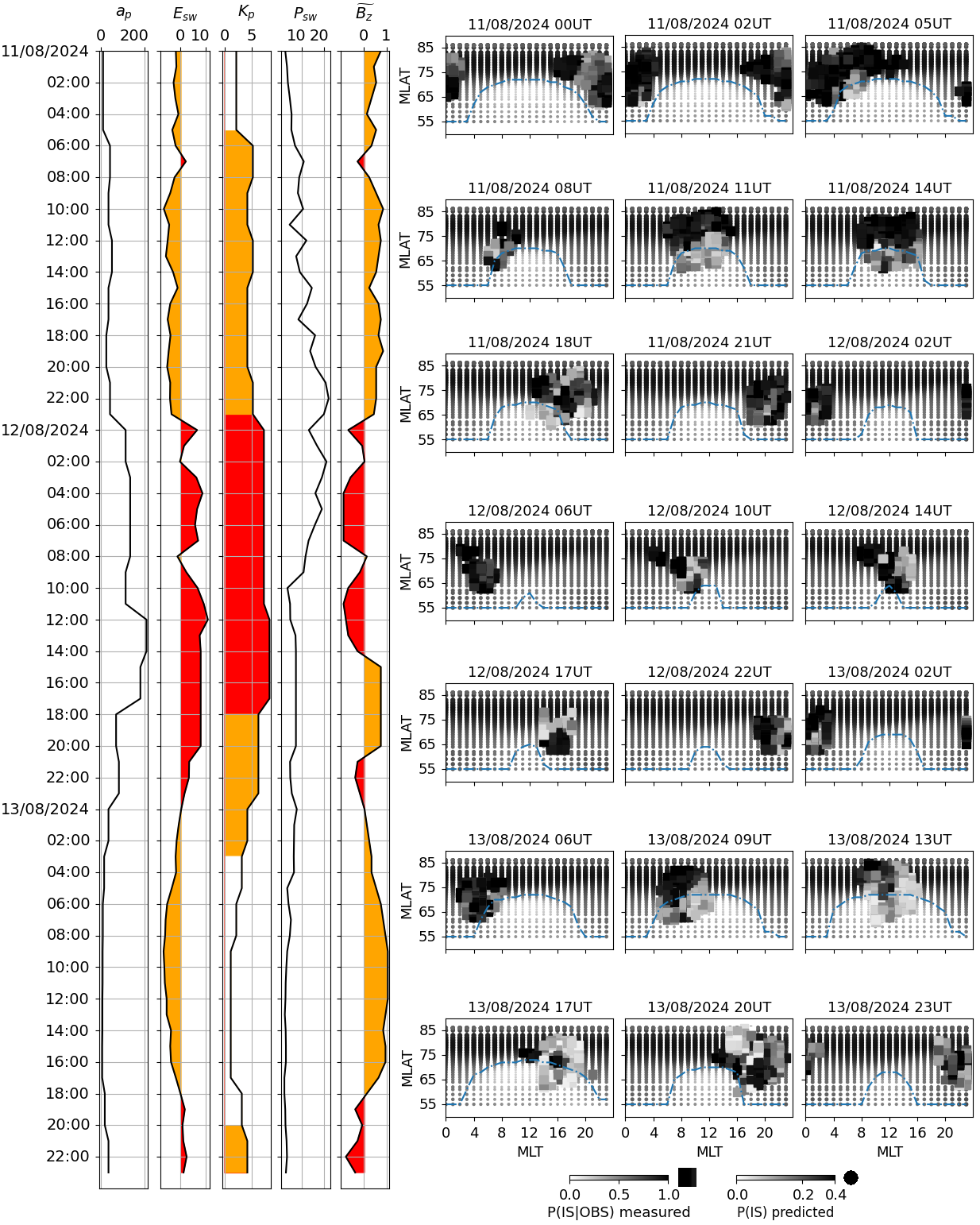}

\caption{Model predictions on the validation dataset in nowcasting mode for
the geomagnetic storm of August 11-13, 2024. The left panel shows
the time series of the indices $a_{p}$, $E_{sw}$, $K_{p}$, $P_{sw}$,
and $\widetilde{B_{z}}$. The right panel displays model-predicted
probabilities of observing ionospheric irregularities $P(IS|\theta_{s})$
as circles, and experimental observations of the probability that
a scattered signal originates from ionospheric scattering $P(IS|OBS,\theta_{s})$
as squares. The line indicates the model border $P(IS|\theta_{s})=0.1$.}
\label{fig:model_P_2023}
\end{figure}

The Fig.\ref{fig:model_P_2023} shows some space weather parameters 
and nowcasting model predictions
using data from the validation dataset - the geomagnetic storm of
August 11-13, 2024. The figure shows good agreement between
experimental observations and model predictions. A characteristic daytime
decrease in the $P(IS|\theta_s)$ is visible near
local noon. It looks like a well-known phenomenon related with the asymmetry
of the polar oval. Figure also
demonstrates a good correlation between the equatorward movement of the irregularities
boundary and increasing the geomagnetic disturbance level.

\begin{figure}
\includegraphics[scale=0.85]{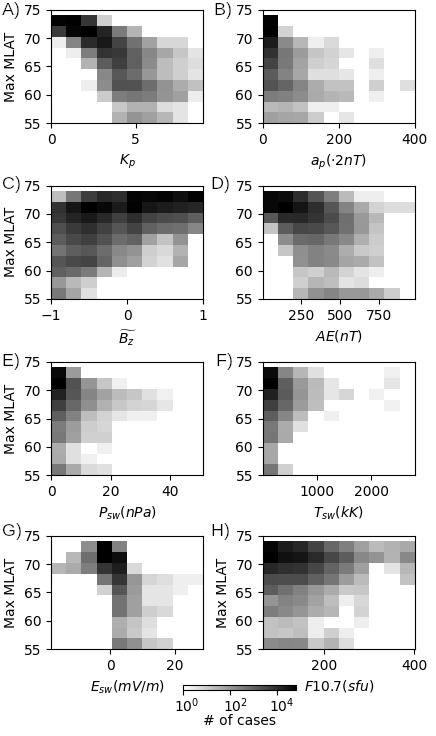}
\caption{Dependence of the maximum latitude of the equatorial boundary of the
model region containing irregularities (nowcasting) on space weather parameters
based on whole available dataset 2010-2025.}
\label{fig:final_corr}
\end{figure}

The dependence of the maximum latitude of the equatorial boundary
of the irregularities (observed in the 10-12 MLT sector) on the space weather parameters,
used as model inputs, is shown in detail in Fig.\ref{fig:final_corr} (for nowcast model).
The statistics shows three types of dependencies affecting the maximum
latitude of the lower boundary of the model region containing irregularities:
\begin{itemize}
\item as the parameters $K_{p}$, $AE$, and $E_{sw}$ increase, the boundary moves
equatorward;
\item as $F10.7$, $T_{sw}$, and $\widetilde{B_{z}}$
increase, the boundary moves poleward;
\item no unambiguous
direct correlation is observed with changes in $P_{sw}$ and $a_{p}$,
the strong dependence on them (Table \ref{tab:fi_forecast_best})
expects that not their realtime values, but their temporal dynamics afftect
the result.
\end{itemize}

The influence of the geomagnetic disturbance level ($K_{p}$) on the
probability of observing small-scale irregularities has long been
known: as the level of geomagnetic disturbance increases,
the probability of observing radio aurora at lower latitudes rises,
which is consistent with the results shown in Fig.\ref{fig:final_corr}A.

The triggering mechanism involving the sign of the interplanetary
magnetic field component $\widetilde{B_{z}}$ (effectively, $\widetilde{B_{z}}$
represents the cosine of IMF clock angle) regarding the probability of a geomagnetic
storm and its associated small-scale irregularities is also well known\cite{Heppner1987}, 
which is consistent with the results shown in Fig.\ref{fig:final_corr}C.

Fig.\ref{fig:final_corr}G demosntrates also strong trigger effect
of $E_{sw}$ - the product of the solar wind velocity magnitude $|V|$
and the z-component of the magnetic field $B_{z}$ with taking its
sign into account.
The dependence of polar convection intensity on the electric field
module $|E_{sw}|$ is utilized in constructing high-latitude convection
maps based on SuperDARN radar data \cite{Cousins_2010}, yet it does
not appear to exhibit a clear, unambiguous relationship with the latitude
of observation. The dependence on $V_{x}$ is also known because it
has been shown to impact the strength of the coupling between the
magnetosphere and ionosphere \cite{Reiff_1981,Doyle_1983}, and is
used in the high-latitude electric potential models \cite{Weimer_1996,Weimer_2005}.
The solar wind pressure $P_{sw}$ has also been shown to have an impact
on convection \cite{Boudouridis_2004}. 

Thus, the empirical model we constructed and the significant parameters
identified within it do not contradict known physical
observations or the parameters controlling substorm development and
high-latitude convection. They are responsible for the formation
of many types of small-scale irregularities in the high-latitude
ionosphere.

\section{Conclusion}

This paper presents an approach to constructing a predictive model
for the probability of occurrence of high-latitude decameter ionospheric
irregularities based on SuperDARN and SECIRA radar data and space
weather parameters. The method is based on self-learned pattern recognition in the
dynamics of space-weather parameters, 
that influence the probability of irregularities formation. The proposed
approach is a data-driven, multiscale method in which the spatial
distribution of irregularities is approximated using a neural network
with an original architecture inspired by graph neural networks
and trained on available experimental data.

The model was trained using data from 11 radars covering the period
2010-2025 (SuperDARN: CVW, ICE, ICW, INV, PGR, PYK, RKN, SAS, STO;
SECIRA: EKB, MAGW). It is shown that the model achieves optimal
performance when using geomagnetic coordinates MLT, MLAT to describe 
irregularities locations instead of geographic coordinates. 
The optimal set of parameters for the best
forecast is a combination of geomagnetic and solar disturbance indices
($a_{p}$, $K_{p}$, $AE$, F10.7) from ground-based observations,
and satellite-measured solar wind and interplanetary magnetic field
parameters ($P_{sw}$, $E_{sw}$, $\widetilde{B_{z}}$, $T_{sw}$).
Hourly mean values for all the parameters were used in the analysis.
The results indicate that acceptable forecast quality is achieved
with a lead time of up to 7-9 hours, while the highest quality is obtained
for nowcast (zero lead time). It is shown that the parameters $\widetilde{B_{z}}$
and F10.7 are significant for nowcasting only and can be excluded for other
lead times. AE and $T_{sw}$ are significant only at lead times above 2 hours and 
can be neglected for short-term predictions.

Statistical relationships between the maximum latitude of the equatorial
boundary of the irregularities and parameters of the model are also
presented, demonstrating a good correlation with previously known effects.

\section*{Asknowledgements}

The work of OB and IL was financially supported by the Ministry of Science and Higher Education.
The results if SECIRA radars were obtained using the Core Shared Research Facility "Angara" 
of ISTP SB RAS (https://ckp-rf.ru/catalog/ckp/3056/). 
The authors acknowledge the use of SuperDARN data. 
SuperDARN is a collection of radars funded by national scientific funding agencies of Australia, 
Canada, China, France, Italy, Japan, Norway, South Africa, United Kingdom, and the United States of America. 
The Christmas Valley West, Iceland East, and Iceland West radars are maintained and operated by 
Dartmouth College with support from NSF under AGS-2426199. 
The Inuvik, Prince George, Rankin Inlet, and Saskatoon radars are 
maintained and operated by SuperDARN Canada. 
SuperDARN Canada operations are supported by funding from the Canada Foundation for Innovation (CFI), 
the Canadian Space Agency's (CSA) Geospace Observatory (GO) Canada program, and Innovation Saskatchewan. 
The Pykkvibaer radar was maintained and operated by University of Leicester. 
The Stokkseyri radar was maintained and operated by  Centre National de Recherche Scientifique (CNRS, France) 
and by Lancaster University (UK).
The OMNI data were obtained from the GSFC/SPDF OMNIWeb interface at
https://omniweb.gsfc.nasa.gov

\bibliographystyle{apalike}
\bibliography{papers4review}

@article{ B_niguel_2011, title={{A global ionosphere scintillation propagation model for equatorial regions}},
	volume={1},
	ISSN={2115-7251},
	DOI={10.1051/swsc/2011004},
	number={1},
	journal={Journal of Space Weather and Space Climate},
	publisher={EDP Sciences},
	author={B\'eniguel, Yannick and Hamel, Pierrick},
	year={2011},
	pages={A04} }

@article{ Booker_1938, title={{Scattering of radio waves by the F‐region of the ionosphere}},
	volume={43},
	ISSN={0096-8013},
	DOI={10.1029/te043i003p00249},
	number={3},
	journal={Terrestrial Magnetism and Atmospheric Electricity},
	publisher={American Geophysical Union (AGU)},
	author={Booker, H. G. and Wells, H. W.},
	year={1938},
	month=Sept, pages={249–256} }

@article{ Caton_2004, title={{GPS proxy model for real‐time UHF satellite communications scintillation maps from the Scintillation Network Decision Aid (SCINDA)}},
	volume={39},
	ISSN={1944-799X},
	DOI={10.1029/2002rs002821},
	number={1},
	journal={Radio Science},
	publisher={American Geophysical Union (AGU)},
	author={Caton, R. G. and McNeil, W. J. and Groves, K. M. and Basu, S.},
	year={2004},
	month=Jan }

@article{ Deshpande_2014, 
        title={{Satellite‐beacon Ionospheric‐scintillation Global Model of the upper Atmosphere (SIGMA) I: High‐latitude sensitivity study of the model parameters}},
	volume={119},
	ISSN={2169-9402},
	url={http://dx.doi.org/10.1002/2013JA019699},
	DOI={10.1002/2013ja019699},
	number={5},
	journal={Journal of Geophysical Research: Space Physics},
	publisher={American Geophysical Union (AGU)},
	author={Deshpande, K. B. and Bust, G. S. and Clauer, C. R. and Rino, C. L. and Carrano, C. S.},
	year={2014},
	month=May, pages={4026–4043} }

@article{ Gao_2025, title={{ISNet: Decomposed Dynamic Spatio‐Temporal Neural Network for Ionospheric Scintillation Forecasts}},
	volume={23},
	ISSN={1542-7390},
	url={http://dx.doi.org/10.1029/2024SW004239},
	DOI={10.1029/2024sw004239},
	number={6},
	journal={Space Weather},
	publisher={American Geophysical Union (AGU)},
	author={Gao, Zhixu and Chen, Yanhong and Ao, Xianzhi and Yue, Fulu and Chen, Hong and Deng, Hao and Luo, Bingxian and Wang, Xin and Yuan, Tianjiao},
	year={2025},
	month=May }

@article{ Gao_2026, title={{A Novel Deep Learning Model for High‐Latitude Ionospheric Convection Using SuperDARN Data}},
	volume={24},
	ISSN={1542-7390},
	url={http://dx.doi.org/10.1029/2026SW005037},
	DOI={10.1029/2026sw005037},
	number={7},
	journal={Space Weather},
	publisher={American Geophysical Union (AGU)},
	author={Gao, Bingjie and Liu, Erxiao and Hu, Zejun and Liu, Jianjun and Yang, Ding and Qiao, Lei and Chen, Xuan},
	year={2026},
	month=July }

@article{ Groves_1997, title={{Equatorial scintillation and systems support}},
	volume={32},
	ISSN={1944-799X},
	url={http://dx.doi.org/10.1029/97RS00836},
	DOI={10.1029/97rs00836},
	number={5},
	journal={Radio Science},
	publisher={American Geophysical Union (AGU)},
	author={Groves, K. M. and Basu, S. and Weber, E. J. and Smitham, M. and Kuenzler, H. and Valladares, C. E. and Sheehan, R. and MacKenzie, E. and Secan, J. A. and Ning, P. and McNeill, W. J. and Moonan, D. W. and Kendra, M. J.},
	year={1997},
	month=Sept, pages={2047–2064} }

@article{ Hamel_2014, title={{Kriging method to perform scintillation maps based on measurement and GISM model}},
	volume={49},
	ISSN={1944-799X},
	url={http://dx.doi.org/10.1002/2014RS005470},
	DOI={10.1002/2014rs005470},
	number={9},
	journal={Radio Science},
	publisher={American Geophysical Union (AGU)},
	author={Hamel, Pierrick and Sambou, Diagna Charles and Darces, Muriel and Beniguel, Yannick and H\'elier, Marc},
	year={2014},
	month=Sept, pages={746–752} }

@article{ Huba_2010, title={{Global modeling of equatorial plasma bubbles}},
	volume={37},
	ISSN={1944-8007},
	url={http://dx.doi.org/10.1029/2010GL044281},
	DOI={10.1029/2010gl044281},
	number={17},
	journal={Geophysical Research Letters},
	publisher={American Geophysical Union (AGU)},
	author={Huba, J. D. and Joyce, G.},
	year={2010},
	month=Sept }

@article{ Mestici_2026, title={{Toward a Dynamic Model for Topside Ionospheric Irregularities: Spherical Harmonic Decomposition}},
	volume={24},
	ISSN={1542-7390},
	url={http://dx.doi.org/10.1029/2025SW004421},
	DOI={10.1029/2025sw004421},
	number={3},
	journal={Space Weather},
	publisher={American Geophysical Union (AGU)},
	author={Mestici, S. and De Michelis, P. and Consolini, G. and Pignalberi, A. and Tozzi, R. and Giannattasio, F. and Berilli, F.},
	year={2026},
	month=Feb }

@article{ Molina_2026, title={{An Improved Neural-Network to Estimate the Inputs of Rino’s Ionospheric Scintillation Model}},
	volume={19},
	ISSN={2151-1535},
	url={http://dx.doi.org/10.1109/JSTARS.2025.3625408},
	DOI={10.1109/jstars.2025.3625408},
	journal={IEEE Journal of Selected Topics in Applied Earth Observations and Remote Sensing},
	publisher={Institute of Electrical and Electronics Engineers (IEEE)},
	author={Molina, Carlos and Camps, Adriano},
	year={2026},
	pages={181–189} }

@article{ Priyadarshi_2015, title={{A Review of Ionospheric Scintillation Models}},
	volume={36},
	ISSN={1573-0956},
	url={http://dx.doi.org/10.1007/s10712-015-9319-1},
	DOI={10.1007/s10712-015-9319-1},
	number={2},
	journal={Surveys in Geophysics},
	publisher={Springer Science and Business Media LLC},
	author={Priyadarshi, S.},
	year={2015},
	month=Jan, pages={295–324} }

@article{ Secan_1995, title={{An improved model of equatorial scintillation}},
	volume={30},
	ISSN={1944-799X},
	url={http://dx.doi.org/10.1029/94RS03172},
	DOI={10.1029/94rs03172},
	number={3},
	journal={Radio Science},
	publisher={American Geophysical Union (AGU)},
	author={Secan, J. A. and Bussey, R. M. and Fremouw, E. J. and Basu, Sa.},
	year={1995},
	month=May, pages={607–617} }

@article{ Secan_1997, title={{High‐latitude upgrade to the Wideband ionospheric scintillation model}},
	volume={32},
	ISSN={1944-799X},
	url={http://dx.doi.org/10.1029/97RS00453},
	DOI={10.1029/97rs00453},
	number={4},
	journal={Radio Science},
	publisher={American Geophysical Union (AGU)},
	author={Secan, J. A. and Bussey, R. M. and Fremouw, E. J. and Basu, S.},
	year={1997},
	month=July, pages={1567–1574} }

@article{ Tsunoda_1988, title={{High‐latitude <i>F</i> region irregularities: A review and synthesis}},
	volume={26},
	ISSN={1944-9208},
	url={http://dx.doi.org/10.1029/RG026i004p00719},
	DOI={10.1029/rg026i004p00719},
	number={4},
	journal={Reviews of Geophysics},
	publisher={American Geophysical Union (AGU)},
	author={Tsunoda, Roland T.},
	year={1988},
	month=Nov, pages={719–760} }

@article{ Zhao_2026, title={{A Model for Predicting Ionospheric Scintillation Intensity in High Latitude Region Based on VMD‐CNN‐BiLSTM‐Attention}},
	volume={131},
	ISSN={2169-9402},
	url={http://dx.doi.org/10.1029/2026JA035098},
	DOI={10.1029/2026ja035098},
	number={6},
	journal={Journal of Geophysical Research: Space Physics},
	publisher={American Geophysical Union (AGU)},
	author={Zhao, Dongsheng and Yang, Dongwei and Liu, Hao and Zhang, Kefei and Hancock, Craig M. and Zhang, Minghao},
	year={2026},
	month=May }

@article{Ruohoniemi_1996,
author = {Ruohoniemi, J. M. and Greenwald, R. A.},
title = {Statistical patterns of high-latitude convection obtained from Goose Bay HF radar observations},
journal = {Journal of Geophysical Research: Space Physics},
volume = {101},
number = {A10},
pages = {21743-21763},
DOI= {10.1029/96JA01584},
url = {https://agupubs.onlinelibrary.wiley.com/doi/abs/10.1029/96JA01584},
eprint = {https://agupubs.onlinelibrary.wiley.com/doi/pdf/10.1029/96JA01584},
year = {1996}
}

@article{McGranaghan_2018,
author = {McGranaghan, Ryan M. and Mannucci, Anthony J. and Wilson, Brian and Mattmann, Chris A and Chadwick, Richard},
title = {New Capabilities for Prediction of High-Latitude Ionospheric Scintillation: A Novel Approach With Machine Learning},
journal = {Space Weather},
volume = {16},
number = {11},
pages = {1817-1846},
DOI= {10.1029/2018SW002018},
url = {https://agupubs.onlinelibrary.wiley.com/doi/abs/10.1029/2018SW002018},
eprint = {https://agupubs.onlinelibrary.wiley.com/doi/pdf/10.1029/2018SW002018},
year = {2018}
}

@article{ Boudouridis_2004, title={{Evaluation of the Hill‐Siscoe transpolar potential saturation model during a solar wind dynamic pressure pulse}},
	volume={31},
	ISSN={1944-8007},
	url={http://dx.doi.org/10.1029/2004GL021252},
	DOI={10.1029/2004gl021252},
	number={23},
	journal={Geophysical Research Letters},
	publisher={American Geophysical Union (AGU)},
	author={Boudouridis, A. and Zesta, E. and Lyons, L. R. and Anderson, P. C.},
	year={2004},
	month=Dec }

@article{ Cousins_2010, title={{A dynamical model of high‐latitude convection derived from SuperDARN plasma drift measurements}},
	volume={115},
	ISSN={0148-0227},
	url={http://dx.doi.org/10.1029/2010JA016017},
	DOI={10.1029/2010ja016017},
	number={A12},
	journal={Journal of Geophysical Research: Space Physics},
	publisher={American Geophysical Union (AGU)},
	author={Cousins, E. D. P. and Shepherd, S. G.},
	year={2010},
	month=Dec }

@article{ Doyle_1983, title={{S3‐2 measurements of the polar cap potential}},
	volume={88},
	ISSN={0148-0227},
	url={http://dx.doi.org/10.1029/JA088iA11p09125},
	DOI={10.1029/ja088ia11p09125},
	number={A11},
	journal={Journal of Geophysical Research: Space Physics},
	publisher={American Geophysical Union (AGU)},
	author={Doyle, M. A. and Burke, W. J.},
	year={1983},
	month=Nov, pages={9125–9133} }

@article{ Reiff_1981, title={{Dependence of polar cap potential drop on interplanetary parameters}},
	volume={86},
	ISSN={0148-0227},
	url={http://dx.doi.org/10.1029/JA086iA09p07639},
	DOI={10.1029/ja086ia09p07639},
	number={A9},
	journal={Journal of Geophysical Research: Space Physics},
	publisher={American Geophysical Union (AGU)},
	author={Reiff, P. H. and Spiro, R. W. and Hill, T. W.},
	year={1981},
	month=Sept, pages={7639–7648} }

@article{ Weimer_1996, title={{A flexible, IMF dependent model of high‐latitude electric potentials having “Space Weather” applications}},
	volume={23},
	ISSN={1944-8007},
	url={http://dx.doi.org/10.1029/96GL02255},
	DOI={10.1029/96gl02255},
	number={18},
	journal={Geophysical Research Letters},
	publisher={American Geophysical Union (AGU)},
	author={Weimer, D. R.},
	year={1996},
	month=Sept, pages={2549–2552} }

@article{ Weimer_2005, title={{Improved ionospheric electrodynamic models and application to calculating Joule heating rates}},
	volume={110},
	ISSN={0148-0227},
	url={http://dx.doi.org/10.1029/2004JA010884},
	DOI={10.1029/2004ja010884},
	number={A5},
	journal={Journal of Geophysical Research: Space Physics},
	publisher={American Geophysical Union (AGU)},
	author={Weimer, D. R.},
	year={2005},
	month=May }

@article{Shepherd_2000,
author = {Shepherd, S. G. and Ruohoniemi, J. M.},
title = {{Electrostatic potential patterns in the high-latitude ionosphere constrained by SuperDARN measurements}},
journal = {Journal of Geophysical Research: Space Physics},
volume = {105},
number = {A10},
pages = {23005-23014},
DOI= {10.1029/2000JA000171},
url = {https://agupubs.onlinelibrary.wiley.com/doi/abs/10.1029/2000JA000171},
eprint = {https://agupubs.onlinelibrary.wiley.com/doi/pdf/10.1029/2000JA000171},
year = {2000}
}

@article{Blanchard_2011,
author = {Blanchard, G. T. and Sundeen, S. and Baker, K. B.},
title = {{Probabilistic identification of high-frequency radar backscatter from the ground and ionosphere based on spectral characteristics}},
journal = {Radio Science},
volume = {44},
number = {5},
pages = {},
DOI = {10.1029/2009RS004141},
url = {https://agupubs.onlinelibrary.wiley.com/doi/abs/10.1029/2009RS004141},
eprint = {https://agupubs.onlinelibrary.wiley.com/doi/pdf/10.1029/2009RS004141},
year = {2009}
}

@article{BERNGARDT_2026,
title = {{Self-learning signal classifier for HF coherent scatter radars}},
journal = {Advances in Space Research},
volume = {77},
number = {3},
pages = {3527-3548},
year = {2026},
issn = {0273-1177},
DOI = {10.1016/j.asr.2025.11.074},
url = {https://www.sciencedirect.com/science/article/pii/S0273117725013584},
author = {Oleg Berngardt and Ivan Lavygin}
}

@misc{WeatherNext3,
      title={WeatherNext 3: Increasing resolution and performance of global weather models with raw observations}, 
      author={Stephan Rasp and Boris Babenko and Dominic Masters and Andrew El-Kadi and Samier Merchant and Guy Shalev and Ilan Price and Fred Zyda and Remi Lam and Sasha Shysheya and Matthew Willson and Stratis Markou and Shreya Agrawal and Suhani Vora and Mohammed Alewi Hassen and Sunny Mak and Tom R. Andersson and Megan Bela and Akib Uddin and Nofar Peled Levi and Ben Gaiarin and Ferran Alet and Aaron Bell and Peter Battaglia and Alvaro Sanchez-Gonzalez},
      year={2026},
      eprint={2609.03582},
      archivePrefix={arXiv},
      primaryClass={cs.LG},
      url={https://arxiv.org/abs/2609.03582}, 
}

@ARTICLE{GATv2,
       author = {{Brody}, Shaked and {Alon}, Uri and {Yahav}, Eran},
        title = {{How Attentive are Graph Attention Networks?}},
      journal = {arXiv e-prints},
         year = 2021,
        month = may,
          eid = {arXiv:2105.14491},
        pages = {arXiv:2105.14491},
          doi = {10.48550/arXiv.2105.14491},
archivePrefix = {arXiv},
       eprint = {2105.14491},
 primaryClass = {cs.LG},
       adsurl = {https://ui.adsabs.harvard.edu/abs/2021arXiv210514491B}
}

@misc{Attentionneed2023,
      title={{Attention Is All You Need}}, 
      author={Ashish Vaswani and Noam Shazeer and Niki Parmar and Jakob Uszkoreit and Llion Jones and Aidan N. Gomez and Lukasz Kaiser and Illia Polosukhin},
      year={2023},
      eprint={1706.03762},
      archivePrefix={arXiv},
      primaryClass={cs.CL},
      url={https://arxiv.org/abs/1706.03762}, 
}

@misc{OMNIdatabase,
      title={{OMNI database}}, 
      author={{OMNI database}},
      year={2026},
      url={https://spdf.gsfc.nasa.gov/pub/data/omni/high\_res\_omni/}, 
}

@misc{HelmholtzCentreForGeosciences,
      title={{Helmholtz Centre For Geosciences}}, 
      author={{Helmholtz Centre For Geosciences}},
      year={2026},
      url={https://kp.gfz.de/en/data}, 
}

@misc{KyotoWDC,
      title={{Kyoto WDC}}, 
      author={{Kyoto WDC}},
      year={2026},
      url={https://wdc.kugi.kyoto-u.ac.jp/aeasy/}, 
}

@article{Nishitani2019,
  title = {Review of the accomplishments of mid-latitude Super Dual Auroral Radar Network (SuperDARN) HF radars},
  author = {Nishitani, Nozomu and Ruohoniemi, John Michael and Lester, Mark and Baker, Joseph Benjamin Harold and Koustov, Alexandre Vasilyevich and Shepherd, Simon G. and Chisham, Gareth and Hori, Tomoaki and Thomas, Evan G. and Makarevich, Roman A. and Marchaudon, Aur{\'e}lie and Ponomarenko, P. and Wild, J. A. and Milan, S. E. and Bristow, W. A. and Devlin, J. and Miller, E. and Greenwald, R. A. and Ogawa, T.},
  journal = {Progress in Earth and Planetary Science},
  volume = {6},
  number = {1},
  pages = {27},
  year = {2019},
  publisher = {Springer},
  doi = {10.1186/s40645-019-0270-5}
}

@article{CNDARN, 
title={{Preliminary Analysis of Observation Data by High Frequency Radars of the CN-DARN}}, 
volume={45}, 
ISSN={0254-6124}, 
DOI={10.11728/cjss2025.02.2024-0057}, 
number={2}, 
journal={Chinese Journal of Space Science}, 
publisher={National Space Science Center, Chinese Academy of Sciences}, 
author={Li, Hang and Zhang, Jiaojiao and Wang, Wei and Deng, Xiang and Lan, Ailan and Zhang, Runzhi and Yan, Jingye and Wang, Chi}, 
year={2025}, 
pages={266} 
}

@article{SECIRA,
  author = {Berngardt, O. I. and Kurkin, V. I. and Kushnarev, D. S. and Grkovich, K. V. and Fedorov, R. R. and Orlov, A. I. and Harchenko, V. V.},
  title = {{ISTP SB RAS decameter radars}},
  journal = {Solar-Terrestrial Physics},
  volume = {6},
  number = {2},
  pages = {63--73},
  year = {2020},
  doi = {10.12737/stp-62202006}
}

@article{IRI2020,
author = {Bilitza, Dieter and Pezzopane, Michael and Truhlik, Vladimir and Altadill, David and Reinisch, Bodo W. and Pignalberi, Alessio},
title = {{The International Reference Ionosphere Model: A Review and Description of an Ionospheric Benchmark}},
journal = {Reviews of Geophysics},
volume = {60},
number = {4},
pages = {e2022RG000792},
doi = {https://doi.org/10.1029/2022RG000792},
note = {e2022RG000792 2022RG000792},
year = {2022}
}

@article{AACGMv2,
author = {Shepherd, S. G.},
title = {{Altitude-adjusted corrected geomagnetic coordinates: Definition and functional approximations}},
journal = {Journal of Geophysical Research: Space Physics},
volume = {119},
number = {9},
pages = {7501-7521},
doi = {https://doi.org/10.1002/2014JA020264},
year = {2014}
}

@article{PermutationFI, 
 volume={45}, 
 ISSN={0885-6125}, 
 DOI={10.1023/a:1010933404324}, 
 number={1}, 
 journal={Machine Learning}, 
 publisher={Springer Science and Business Media LLC}, 
 author={Breiman, Leo}, 
 year={2001}, 
 pages={5-32} 
}

@article{GNN, 
title={{A review of graph neural networks: concepts, architectures, techniques, challenges, datasets, applications, and future directions}}, 
volume={11}, 
ISSN={2196-1115}, 
url={http://dx.doi.org/10.1186/s40537-023-00876-4}, 
DOI={10.1186/s40537-023-00876-4}, 
number={1}, 
journal={Journal of Big Data}, 
publisher={Springer Science and Business Media LLC}, 
author={Khemani, Bharti and Patil, Shruti and Kotecha, Ketan and Tanwar, Sudeep}, 
year={2024}, 
month=Jan 
}

@article{Heppner1987,
author = {Heppner, J. P. and Maynard, N. C.},
title = {Empirical high-latitude electric field models},
journal = {Journal of Geophysical Research: Space Physics},
volume = {92},
number = {A5},
pages = {4467-4489},
doi = {https://doi.org/10.1029/JA092iA05p04467},
year = {1987}
}

\end{document}